\documentclass[10pt,preprint]{aastex}

\newcommand{\teff}{T$_{\rm eff}$}
\newcommand{\logg}{log(g)}
\newcommand{\feh}{\rm [Fe/H]}

\slugcomment{in press at ApJ}
\shortauthors{Everett et al.}
\shorttitle{Spectra of {\it Kepler} Exoplanet Candidate Host
  Stars}

\begin{document}

\title{
Spectroscopy of Faint {\it Kepler} Mission
Exoplanet Candidate Host Stars
}

\author{
Mark E. Everett \altaffilmark{1},
Steve B. Howell \altaffilmark{2,4},
David R. Silva \altaffilmark{1},
and Paula Szkody \altaffilmark{3,4}
}

\altaffiltext{1}{National Optical Astronomy Observatory, 950 N. Cherry
  Ave, Tucson, AZ 85719, USA} 
\altaffiltext{2}{NASA Ames Research Center, Moffett Field, CA 94035,
  USA} 
\altaffiltext{3}{Dept. of Astronomy, University of Washington,
  Seattle, WA 98195, USA}
\altaffiltext{4}{Visiting Astronomer, Kitt Peak National Observatory,
  National Optical Astronomy Observatory, which is operated by the
  Association of Universities for Research in Astronomy (AURA) under
  cooperative agreement with the National Science Foundation.}

\setcounter{footnote}{4} 

\begin{abstract}
Stellar properties are measured for a large set of {\it Kepler}
Mission exoplanet candidate host stars.  Most of these stars are
fainter than ${\rm 14^{th}}$ magnitude, in contrast to other
spectroscopic follow-up studies.  This sample includes many
high-priority Earth-sized candidate planets.  A set of model spectra
are fitted to $R\sim3000$ optical spectra of 268 stars to improve
estimates of \teff, \logg, and \feh\ for the dwarfs in the range ${\rm
  4750\leq T_{eff} \leq7200}$~K.  These stellar properties are used to
find new stellar radii and, in turn, new radius estimates for the
candidate planets.  The result of improved stellar characteristics is
a more accurate representation of this {\it Kepler} exoplanet sample
and identification of promising candidates for more detailed study.
This stellar sample, particularly among stars with \teff$\gtrsim5200$~K,
includes a greater number of relatively evolved stars with larger
radii than assumed by the mission on the basis of multi-color
broadband photometry.  About 26\% of the modelled stars require radii
to be revised upwards by a factor of 1.35 or greater, and modelling of
87\% of the stars suggest some increase in radius.  The sample
presented here also exhibits a change in the incidence of planets
larger than ${\rm 3-4R_{\oplus}}$ as a function of metallicity.  Once
\feh\ increases to $\geq{-0.05}$, large planets suddenly appear in the
sample while smaller planets are found orbiting stars with a wider
range of metallicity.  The modelled stellar spectra, as well as an
additional 84 stars of mostly lower effective temperatures, are made
available to the community.
\end{abstract}

\keywords{
planetary systems -- planets and satellites: fundamental parameters --
stars: fundamental parameters -- surveys
}

\section{INTRODUCTION}\label{sect:introduction}
The NASA {\it Kepler} Mission employs a space-based 0.95~m aperture
Schmidt telescope to observe a single 115 square degree field of view
and obtain nearly continuous light curve coverage for over 156,000
stars.  The satellite was launched in March 2009 and began science
observations in May 2009 with a primary mission objective of detecting
the transits of small planets orbiting near the habitable zone of
Sun-like stars \citep{boruckietal10}.

Once detrended for instrumental signatures and long-term stellar
variations, the {\it Kepler} light curves are searched for transit
signals that are vetted to eliminate likely false positives
\citep[transit-like signals due to causes other than transiting
  planets; see][]{batalhaetal10}.  The periodicity and amplitude of
the transits provide initial estimates for orbital periods and sizes
of candidate planets.  However, these planet size estimates are
derived from modeling the light curves with a parameter reflecting the
planet-to-star radius ratio and so depend on the uncertainty of the
radius of the host star.  Understanding the properties of the host
stars, especially stellar radii, is therefore critical to meeting many
of the mission objectives.  In order to identify the most promising
candidates, refine knowledge of the host star properties, and identify
additional false positives, a follow-up observing program was
undertaken to obtain optical spectra of candidate host stars.  The
resulting spectra are fitted with models to determine the three
stellar properties \teff, \logg, and \feh.\ These parameters are then
used to revise the stellar and candidate planet radii.  This program
is one of several providing ground-based follow-up reconnaissance
spectroscopy of candidate exoplanet host stars as part of the {\it
  Kepler} Follow-up Program \citep{gautieretal10}.

The target sample is described in \S\ref{sect:sample}, the
observational methods in \S\ref{sect:observations}, and the data
reduction in \S\ref{sect:datareduction}.  In
\S\ref{sect:characterization} model fits are used to determine the
stellar properties \teff, \logg, and \feh\ along with an analysis of
their uncertainties.  These stellar parameters are used in
\S\ref{sect:radii} to find fits for each star on sets of isochrones
and derive revised stellar and planetary radii.  The results are
discussed in \S\ref{sect:discussion} and presented in a table listing
the stellar properties for 220 candidate exoplanet host stars.  The
public availability of the data are discussed in
\S\ref{sect:dataavailable} and the findings from these data are
summarized in \S\ref{sect:conclusions}.

\section{TARGET SAMPLE}\label{sect:sample}

The target stars are selected from a list of candidate exoplanet host
stars known as {\it Kepler} Objects of Interest (KOIs) identified by
the mission following a battery of tests that is designed to identify
false positives.  These tests include a manual inspection of each
light curve and analysis of any pixel-level flux and centroid
variations during the candidate transits \citep{batalhaetal10}.
Having passed the initial false positive identification tests
unscathed, KOIs can be considered reasonable targets for planet
characterization and confirmation as bona-fide planets using
ground-based follow-up observations.  At this point, the KOI list
contains some unidentified false positives with a rate
that depends on the system's properties.  Theoretical calculations
have been used to predict the rate of false positives due to eclipsing
binaries, especially cases where flux of a third star is blended with
the eclipsing binary.  \citet{mortonjohnson11} predicted an overall
false positive rate of 5\% based on galactic structure models, the
expected binary star population and eclipse depths.  Later,
\citet{morton12} pointed out that because the KOI list still contains
some candidates with V-shaped light curves, a higher false positive
rate might be expected.  \citet{fressinetal13} carried out a recent
analysis that included simulating eclipsing binaries as background
sources or as members of heirarchical triple systems and systems where
true planets had their light curves blended with the flux of other
stars.  Their analysis predicted a higher overall false positive rate
of 9.4\% with a dependence on the presumed planet radius and galactic
latitude.  The highest false positive rate of 17.7\% was predicted for
giant planet candidates.  Recent observational studies have also
pointed toward a significant false positive rate.
\citet{santerneetal12} conducted a radial velocity survey and
estimated a 35\% false positive rate among short-period giant planet
candidates.  \citet{colonetal12} used multi-color light curves to find
two out of a sample of four short-period small planet KOIs were
actually eclipsing binary stars, necessitating a comparably high false
positive rate.  Stellar classification spectroscopy can identify false
positives in cases where stellar properties are found to be incorrect,
however other types of observations are typically better suited to
identifying individual false positive candidates.

Our spectra were most often the first follow-up observations taken of
the faint stars of interest.  Up to this point, these stars have
normally been characterized based only on modelling of the broadband
photometry contained in the {\it Kepler} Input Catalog, a ground-based
survey of the {\it Kepler} field \citep[KIC;][]{brownetal11}.  The
stellar properties determined in the KIC were designed to select
optimal target stars for the mission prior to launch.  The ideal
target stars were small (ie. dwarfs) for which transits by a given
size planet produce relatively large signals.  The KIC allowed {\it
  Kepler} to select mostly small stars, but within the sample, stars
exhibit a range of properties that are not always accurately
determined.

A list of current active KOIs is maintained by the Community Follow-up
Program (CFOP\footnote{https://cfop.ipac.caltech.edu/}) and is
continuously updated as the {\it Kepler} satellite observations are
reduced and vetted for new candidates, or as follow-up observations
help to identify some KOIs as false positives.  The properties of the
KOIs in our sample have likewise changed over the course of the
mission.  The highest priority targets, and those selected to be
included in the sample, generally fall into one or more of the
following categories: (1) KOIs that are requested for observation as
part of an intensive study of a single star or a small number of host
stars, (2) KOI stars that are candidates to be hosts of small planets
($R_p\lesssim2.5R_{\oplus}$), (3) KOIs in which the candidate planets
orbit in a predicted habitable zone, and (4) KOI stars that harbor
multiple candidate planets.

Because the KOIs are also being pursued by other spectroscopic
follow-up programs and we wish to avoid unnecessary overlap, we have
also selected targets on the basis of apparent brightness.  The target
stars span an apparent brightness range ${\rm 8<m_{Kep}<16}$, where
${\rm m_{Kep}}$ is the {\it Kepler} bandpass magnitude
\citep{brownetal11}.  Figure~\ref{Fig:kepmags_hist} shows the
magnitude distribution of our target sample along with the current set
of KOI stars.  It also includes the magnitude distribution of those
stars with new stellar radius estimates (see \S\ref{sect:radii}).  At
the bright end of the magnitude range, ${\rm m_{Kep}<13}$, the
follow-up coverage by other groups is fairly extensive (the {\it
  Kepler} Follow-up Program reported approximately 90\% of these stars
as having had spectral follow-up through the 2012 observing season).
To our knowledge, our spectroscopic sample is by far the largest for
candidate host stars of ${\rm m_{Kep}>14}$ ($N=305$).  Such faint
stars may prove too difficult or time consuming for other follow-up
methods (e.g., highly precise radial velocity measurements), however
they are quite important to the overall mission goals due to their
large numbers (ie. two thirds of currently-active KOI stars have ${\rm
  m_{Kep}>14}$ and two thirds of planet candidates with radii less
than ${\rm 2.5R_\oplus}$ occur around these fainter stars).  A full
understanding of {\it Kepler} exoplanet statistics requires large
follow-up studies of the faint stars, or at least those of highest
priority.  Finally, note that a few otherwise high priority targets
are excluded from the observations due to visible crowding by other
stars since the modelling described here is not designed for composite
spectra.

\begin{figure}
\epsscale{0.55}
\plotone{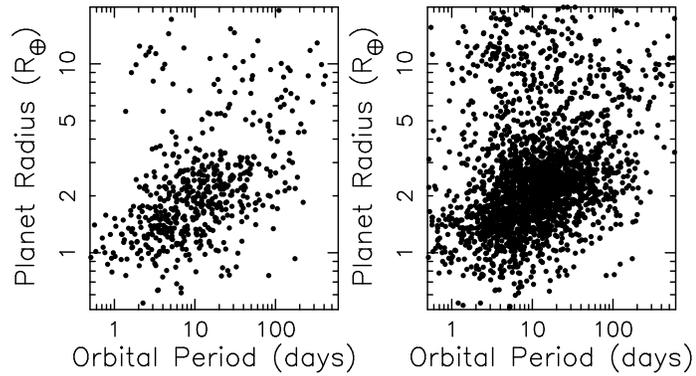}
\caption{The distribution of the initial (pre-follow-up) planet radii
  and orbital periods for the planet candidates hosted by the observed
  stars (left panel) and for those around all KOI stars known as of September
  2012 (right panel).  Both samples are dominated by planets
  smaller than ${\rm 4R_{\oplus}}$, while the observed sample contains
  a higher fraction of small candidate planets and those in
  long-period orbits.  See \S\ref{sect:sample} for a discussion.
} \label{Fig:PpRp}
\end{figure}

Figure~\ref{Fig:PpRp} shows distributions of planet orbital periods
and radii for the same data sets, namely our sample and that of all
KOIs.  The entire KOI sample is dominated by planets smaller than
$4R_{\oplus}$.  As {\it Kepler} obtains increasingly long time
coverage light curves, the relative fractions of small planets and
those in long-period orbits grows and the lower right hand regions in
the plots of Figure~\ref{Fig:PpRp}, where habitable terrestrial
planets may be located, are becoming increasingly well populated.  As
shown in Figure~\ref{Fig:PpRp}, the observed sample has a similar
distribution to the entire KOI sample, but contains relatively fewer
stars harboring large planets and relatively more candidates with
long-period orbits.

\section{OBSERVATIONS}\label{sect:observations}

We observed KOIs in the {\it Kepler} field (115 square degrees
centered at ${\rm \alpha=19^h25^m}$, $\delta=+44.5\arcdeg$) on 48
nights during $2010-2012$ using the National Optical Astronomy
Observatory (NOAO) Mayall 4m telescope on Kitt
Peak and the facility RCSpec long-slit spectrograph with one of its
$2048^2$ pixel CCDs (either T2KA or T2KB).  The spectrograph
configuration was the same on each observing run.  The slit was
1.0$\arcsec$ wide by 49$\arcsec$ long and oriented with a position
angle of 90$\arcdeg$.  The KPC-22b grating in second order was used to
disperse the spectra with 0.72~\AA~pixel$^{-1}$ at a nominal
resolution of $\delta\lambda = 1.7$\AA.  The spectra covered a
wavelength range between 3640\AA\ and 5120\AA, but were out of focus
at both ends where the fluxes could not be reliably calibrated.  The
effective wavelength range was therefore reduced to a
$3800-4900$\AA\ region.  The scale along the slit in each spectrum was
0.69${\rm \arcsec~pixel^{-1}}$.

The observing procedure was basically the same each night.  The
telescope autoguider was used during each observation and each pointing
began with an exposure of the instrument's comparison arc lamp
spectrum (HeNeAr or FeAr) for wavelength calibrations.  Following
that, normally a single exposure was taken of each target star.  The
exposure times ranged between 5 and 20 minutes for most KOIs, although
a few required longer integrations due to faintness or poor observing
conditions.  The faintest targets requiring an exposure time exceeding
20 minutes were observed in two exposures to reduce the density of
cosmic ray hits per exposure and aid in their removal during
reduction.  The KOIs or other stars (e.g., for flux calibrations) were
all observed at an airmass of less than $\sim1.8$.  At the high end of
this airmass range, the atmospheric dispersion for objects in the {\it
  Kepler} field remained sufficiently parallel to the slit at the
latitude of Kitt Peak, and permitted efficient operations at a single
instrument rotation.  At least one spectrophotometric standard star
selected from \citet{masseyetal88} or \citet{stone77} was observed each
night.  Calibration data consisting of bias frames, quartz lamp flat
field exposures, and comparison lamp exposures were taken during the
daytime.

\section{DATA REDUCTION}\label{sect:datareduction}

The data reduction is primarily based on various IRAF\footnote{IRAF is
  distributed by the National Optical Astronomy Observatories, which
  are operated by the Association of Universities for Research in
  Astronomy, Inc., under cooperative agreement with the National
  Science Foundation.} packages for performing image reduction and the
{\it onedspec} package for extracting and calibrating the spectra.
The first step is reducing the sets of bias and quartz flat lamp
exposures.  The overscan bias level is subtracted from each bias frame
and it is trimmed to a useful data section.  These bias frames are
averaged to create a master.  The overscan bias level is subtracted
from each flat field frame followed by any (residual) bias pattern in
the master bias.  The flat frames are then averaged while rejecting
cosmic ray hits.  We normalize this master flat by fitting a smooth
curve to its shape along the dispersion axis (rows) and normalizing
each row of the flat by this curve.  Object spectra frames are reduced
by subtracting the overscan bias, trimming them, and subtracting any
residual bias pattern.  They are then divided by the normalized flat
field.

The onedspec package task {\it doslit} is the basis of spectral
extraction and calibration using the spectrophotometric standard
stars.  To reduce the systematic trends that may result from the
variation in telescope focus with wavelength across the spectrum (a
significant effect with this spectrograph configuration), a relatively
wide aperture is defined to extract each spectrum.  This is based on a
cut through the CCD column at ${\rm 4950\AA}$ where the stellar
profile along the slit is broad and representative of the wavelength
region used for much of the spectral modelling.  We measure sky flux
in regions extracted from both sides of the stellar spectrum, and
subtract it.  The aperture defined by the stellar spectrum is used to
extract a comparison lamp spectrum for each object.  A sensitivity
function is found for each night based on the ratio of the standard
star to its standard curve in the IRAF database of KPNO IRS standards
and used to correct the science targets and supply a relative flux
level.  The comparison lamp spectra are used to determine wavelength
as a function of columns in order to resample the spectra to a linear
wavelength scale set to closely match the sampling of the 2-D spectra.

\section{STELLAR CHARACTERIZATION}\label{sect:characterization}

\subsection{Overview of the Stellar Characterization Methodology}

We developed specialized software and procedures for this program.
These are first used to find the basic stellar properties \teff,
\logg, and \feh, by fitting the observed spectra to theoretical model
spectra (\S\ref{sect:characterization}).  Following that, stellar and
planetary radii are estimated based on the best fits of the basic
stellar properties to Yale-Yonsei isochrones in
(\S\ref{sect:radii}).

The model-fitting methods employed here rely on comparisons between
observed spectra and existing synthetic spectra calculated from
stellar atmosphere models and line modelling codes.  Model spectra are
available from the literature on a grid of discrete values for \teff,
\logg, and \feh.  The process followed here finds the best physical
properties for each observed spectrum by evaluating the rms of the
residuals between the observed spectrum and each model.  Following
that, an interpolated value is found for each stellar property by
evaluating the goodness-of-fit over the grid of models.  Ideally, a
best-fitting model spectrum could be identified for each star and the
physical properties associated with the model would then be assigned
to the star.  In practice, the spectral models fail to accurately
predict all of the features in the observed spectra and fits often
need to be restricted to specific wavelength intervals containing
features that are both well represented by the models and sensitive to
the parameters being sought \citep[e.g., see][]{valentifischer05}.
Furthermore, systematic errors in determining stellar properties can
be introduced by errors in the relative spectral flux calibration and
the discrete values of the model atmosphere sets.  Errors in the
stellar parameters can be correlated as well, complicating the
situation.

In order to test the methods employed here and refine them to work on
KOI stars, we observed a large number of ``test stars'' that have
published stellar properties.  Experimentation has shown that the fits
can reproduce the relative stellar properties for these stars, but
with systematic offsets from their literature values.  The final
fitting methods adopted are ones that best reproduced the literature
values, once corrections for these systematic offsets were made.  In
essence, we adopted the test stars and their published properties as a
standard set and worked to find methods that maximized the fitting
precision.

The model fits are confined to a relatively narrow wavelength region
at the long wavelength end of the spectra (see
\S~\ref{subsec:modelfitting}).  This region contains the important
H$\beta$ absorption line, which is strong in the hotter stars of our
sample.  Its strength and profile is dependent on effective
temperature and surface gravity.  Multiple atomic metal lines are also
present, the strongest ones are due to low ionization states of Fe,
Cr, Mn, Ni, Ti and Mg.  For cooler stars, a prominent broad molecular
feature appears from MgH (near 4780\AA) and eventually from TiO (near
4760\AA) at the lowest temperatures.  These features, and the range of
stellar atmosphere conditions over which they are useful diagnostics,
limit the stars that we can model using these procedures.  During the
fitting, the strength of the metal lines drives our estimate of \feh,
while the strength and profile of H$\beta$ is largely responsible for
driving the fits of \teff\ and \logg.  The strength of MgH is not very
well represented in the synthetic spectra \citep{wecketal03} nor does
this feature appear to be particularly helpful for fitting the cooler
range of our spectra where it appears.  The stars that could be fit
most effectively and for which we had representative test stars were
dwarfs within the effective temperature range ${\rm
  4750K<T_{eff}<7200K}$ (approximate spectral types
K2V through F0V) as discussed in \S\ref{subsec:teststarfits}.
For this reason, only stellar properties for KOIs within this
temperature range are reported here.

\subsection{Model Spectra}\label{subsec:modelspec}

The fits are based on a set of synthetic model spectra made publicly
available by \citet{coelhoetal05}.  These model spectra are calculated
using their extensive line calculation codes along with the model
stellar atmospheres of \citet{castellikurucz03}.  They represent
predictions for non-rotating stars with relative metal abundances
set to the solar values of \citet{grevesseandsauval98}.
The model set includes spectra calculated for
stars that lie at discrete points on a 3-D grid defined by the
parameters \teff, \logg, and \feh.  The model spectra are calculated
at wavelength steps of 0.02\AA\ and with a range and spacing between
adjacent values of each parameter as follows: \teff\ between 3500~K
and 7000~K in steps of 250~K, \logg\ between 1.0 and 5.0 in steps of
0.5, and \feh\ between $-2.5$ and $+0.5$ in steps of 0.5 with an
additional set of models at \feh$=+0.2$.

\subsection{Test Stars}\label{subsec:teststars}

The methods used to fit model spectra to the observations are the
result of experiments fitting models to a set of spectra obtained for
44 test stars while attempting to reproduce the physical parameters
previously published for these stars.  These stars were representative
of the majority of the KOIs we planned to target.  Physical data taken
from the literature for these stars is given in
Table~\ref{tab:teststars} along with the model fitting results
discussed later.  The test stars include a set of 20 exoplanet host
stars characterized by the HATnet project \citep{bakosetal02}.  The
HAT stars are dwarfs ranging in \teff\ between 4591~K and 6600~K,
\logg\ between 4.13 and 4.63, and metallicities between $-0.36$ and
$+0.41$.  Typical uncertainties in their stellar properties are 80~K
for \teff, 0.04 in \logg, and 0.08 for \feh.  The atmospheric
parameters of these stars have been estimated by combining
spectroscopic fits with light curve modelling.  The properties
\teff\ and \feh\ were found using the model atmospheres and line
synthesis code provided by the software Spectroscopy Made Easy
\citep[SME;][]{valentipiskunov96} and \logg\ was found by modelling
the transit light curve parameterized by the ratio of the orbital
semi-major axis to the stellar radius, $a/R_\star$, in the manner of
\citet{sozzettietal07}.  A second set of 6 dwarfs from the work of
\citet{valentifischer05} was observed.  These stars ranged in
\teff\ between 4969~K and 5903~K, \logg\ between 3.97 and 4.85, and
\feh\ between $-1.14$ and $+0.22$ based on SME and \citet{kurucz92}
ATLAS9 atmosphere models.  Another set of 4 dwarfs are KOIs that had
also been observed by other Kepler follow-up programs using
high-resolution spectroscopy (labelled with KOI or Kepler
designations).  These programs utilized SME along with other
constraints.  We also observed a number of evolved stars, including a
set of 5 giants in the Kepler field for which properties have been
derived from astroseismological analysis \citep{kallingeretal10} with
updated results as determined in \citet{kallingeretal12}.  These stars
ranged in \teff\ between 4153~K and 4893~K, \logg\ between 1.66 and
3.27, and \feh\ between $-0.29$ and $+0.18$.  A set of 8 bright giants
from \citet{luckheiter07} was included.  The stellar properties
adopted for this work were those \citet{luckheiter07} derived
spectroscopically using \ion{Fe}{1} and \ion{Fe}{2} lines.  Their
spectral line fits were based on MARCS \citep{gustafssonetal03}
atmosphere models and a variant of the MOOG line synthesis code
\citep{sneden73}.  These stars had properties determined from
high-resolution spectroscopy and spanned a \teff\ range from 4605~K to
7000~K, a \logg\ range from 2.49 to 3.31, and a \feh\ range from
$-0.52$ to $+0.31$.  In addition to the giants, a single dwarf from
\citet{luckheiter07} is included.

\subsection{Model-Fitting Method}\label{subsec:modelfitting}

To determine stellar properties for both our test stars and KOI stars,
we apply an iterative method of fitting model spectra to our
observations, finding one stellar atmosphere parameter at a time, and
in many cases holding other parameters at fixed values until the
best-fitting set of stellar properties is identified.  First we
describe the basic procedures common to every fitting iteration, and
then follow that with the details of each iteration.

To prepare the model spectra, the model data of \citet{coelhoetal05}
are re-binned at a wavelength sampling of 0.3\AA\ for calculation
speed.  Then, for each observed spectrum, the models are resampled
onto the wavelength scale of the observed spectra and smoothed using a
Gaussian kernel with a FWHM of 1.5\AA\ to match the observed
resolution.  Next, based on experiment, a specific wavelength interval
is chosen for each fitting step.  The first procedure during each
iteration is to find the cross-correlation function between the
observed and model spectra, where the mean fluxes of both spectra have
been subtracted.  We use the location of the cross-correlation
function peak to shift the model spectra
to match the observations (correcting for any wavelength
calibration errors and, to first order, any Doppler shift).  Next,
with the mean flux (F$_\lambda$) of our observed spectrum normalized
(but with no normalization relative to a continuum flux done), we
scale the flux of each model spectrum to minimize the rms residuals of
the fit.  This scaling is done with either one or two free parameters:

\begin{equation}\label{eqn:AA}
{\rm F_{scaled} = F_{\lambda}(A + B\lambda)} 
\end{equation}

\noindent
Here, ${\rm F_{\lambda}}$ represents the model flux, the parameter A
represents a simple scaling factor, and B an additional term that
corrects slope differences between the observations and model.  During
some iterations, B is fixed at zero.  The parameter B proved to be
useful in our tests and probably removes systematic errors that might
otherwise adversely affect the fits.

\begin{figure}
\epsscale{0.35}
\plotone{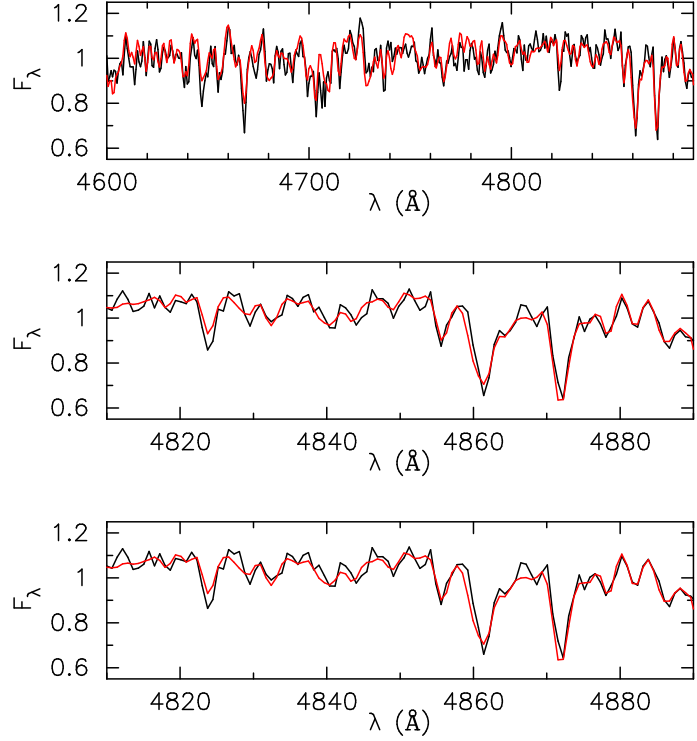}
\caption{The observed spectrum of KOI~2931 in normalized flux units
  (in black) and model atmosphere spectrum (in red) at three steps of
  our model-fitting process as outlined in
  \S~\ref{sect:characterization}.  The top panel shows the fit to find
  \feh$=0.0$ (resulting from step 3 in our process), the middle panel
  shows the fit to find \teff$=5000$~K (resulting from step 4 in our
  process), and the bottom panel shows the fit to find \logg$=4.5$
  (resulting from step 5 in our process).
} \label{Fig:K02931spectra}
\end{figure}

\begin{figure}
\epsscale{0.55}
\plotone{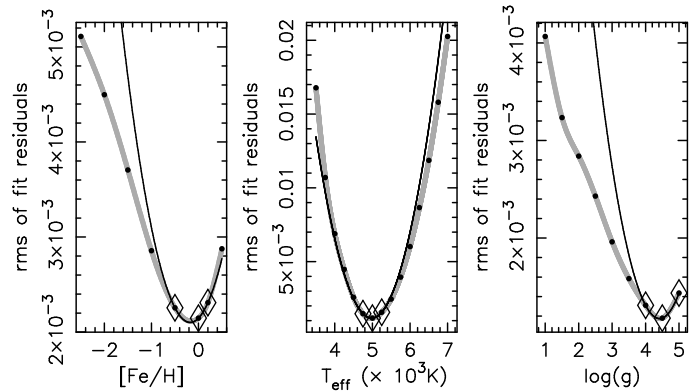}
\caption{The rms of the fitting residuals (filled circles) for model
  spectra fits to our observations of KOI~2931 as a function of three
  stellar parameters and as described in \S~\ref{subsec:modelfitting}.
  The fits shown are for sets of models that vary in one parameter
  with the other two parameters fixed at their values for our
  best-fitting model of \teff$=5000$~K, \logg$=4.5$, and \feh$=0.0$.
  The left panel shows fits as a function of \feh, the middle panel
  shows fits as a function of \teff, and the right panel shows fits as
  a function of \logg.  The grey line shows a cubic spline fit to the
  points.  Open diamond symbols indicate three points selected to
  define our best parameter fit around the minimum rms value.  The
  black lines are quadratic fits through these points whose minima
  represent our preliminary interpolated stellar parameters of
  \teff$=4985$~K, \logg$=4.42$, and \feh$=+0.17$.  These values are
  later corrected for systematic trends as discussed in
  \S~\ref{subsec:teststarfits}.
} \label{Fig:K02931interp}
\end{figure}

We apply the aforementioned procedures in the following step-by-step
process:
\begin{enumerate}
\item An initial value for \feh\ is found by fitting over
  $\lambda=4600-4890$\AA\ for the full set of models while including B
  as a free parameter.  The value of \feh\ for the model having the
  minimum rms of fitting residuals is taken as an initial estimate.
\item An initial value of \teff\ is found by restricting our fits to
  models with \feh\ equal to that found in step 1.  Here, the fit is
  done over the wavelength interval $\lambda=4810-4890$\AA\ and B is
  fixed at zero.  The value of \teff\ for the model having the minimum
  rms of the fitting residuals is taken as an initial estimate.
\item The spectrum is refit to find \feh\ in the manner of step 1, but
  this time the model set is restricted to include only those models
  having \teff\ equal to that found in step 2.  See
  Figure~\ref{Fig:K02931spectra} (top panel) for an example fit to the
  spectrum of KOI~2931 where \feh\ is determined to be $0.0$~dex
  during an application of this step.
\item The spectrum is refit to find \teff\ in the manner of step 2,
  but this time the model set is restricted to include only those
  models having \feh\ equal to that found by step 3.  See
  Figure~\ref{Fig:K02931spectra} (middle panel) for an example fit to
  the spectrum of KOI~2931 where \teff\ is determined to be $5000$~K
  during an application of this step.
\item The value of \logg\ is determined while holding fixed the values
  of \feh\ and \teff\ at the values found in steps 3 and 4
  respectively.  This best-fitting model represents the best gridpoint
  fit to the observed spectrum.  See Figure~\ref{Fig:K02931spectra}
  (bottom panel) for an example fit to find ${\rm log(g)=4.5}$ for
  KOI~2931 during the application of this step.
\item Finally, an interpolated value for each parameter is found as
  described below and illustrated in Figure~\ref{Fig:K02931interp} for
  the case of KOI~2931.  First, each parameter is fitted in turn,
  keeping the values for the two parameters not being fit fixed to
  match their values in the model found in step 5.  The set of models
  fit is thus a function of a single parameter.  The rms values of the
  fitting residuals for these models are considered as a function of
  the parameter value.  To find a minimum over a continuous
  distribution of the parameter value, a cubic spline is fit through
  these data points to locate the minimum.  Then a set of $3-4$ points
  is selected surrounding this minimum and a quadratic function is fit
  through them.  The minimum of the quadratic function is taken to be
  the interpolated parameter value.  In cases where the minimum lies
  at the edge of the grid of parameter values, no
  interpolation can be done.  The spectra in such cases are noted and
  their fits are treated with extra caution.
\end{enumerate}

\subsection{Calibrations Using Test Star Fits}\label{subsec:teststarfits}

As mentioned previously, the stellar properties derived from model
fits such as those performed here are subject to systematic errors
that are difficult to resolve.  Instead of finding a fitting method
free of such errors (which might not be possible), we have chosen to
calibrate these errors based on fits to a set of test star spectra
with the previously-published spectral properties described in
\S\ref{subsec:teststars}.  Once the systematic errors are properly
calibrated, a post-fitting correction is possible.  At the same time,
this approach permits an estimate of the uncertainties in the final
stellar properties.

The results of the model fits to the 44 test star spectra are given in
Table~\ref{tab:teststars}.  For each star, the previously-published
properties are listed with their uncertainties.  Following those are
the values from the fits to our spectra (not yet corrected for the
systematic errors we are attempting to quantify here).  The columns
under the heading ``difference in values'' list the value for each
parameter from this work minus the previously-published value.  To
make the comparison, we plot the difference between our measured
parameter and those from the other literature as a function of our
parameter values.  The results are shown in
Figures~\ref{Fig:feh_trend}, \ref{Fig:teff_trend}, and
\ref{Fig:logg_trend} for \feh, \teff, and \logg\ respectively.  In
each figure, the error bars represent the uncertainties quoted for the
previously-published values.  Note that in these figures, only some of
the test star data are shown, namely a subset of 24 spectral fits that
satisfy one or both of the following restrictions expressed in terms
of the interpolated parameter values obtained during step 6 of the
model fitting procedure:

\begin{equation}\label{eqn:BB}
{\rm
  4761K{\leq}~T_{eff}\leq6998K~\cap~log(g)\geq3.44~\cap~[Fe/H]\geq-0.62}
\end{equation}

\begin{equation}\label{eqn:CC}
{\rm
  5446K{\leq}~T_{eff}\leq6998K~\cap~log(g)\geq3.01~\cap~[Fe/H]\geq-0.62}
\end{equation}

\noindent
The values of the parameter limits in equations~\ref{eqn:BB} and
\ref{eqn:CC} are chosen specifically so that after applying the
corrections for systematic errors the same limits are expressed using
convenient parameter values as discussed below.  The reason that 20 of
the 44 test star spectra are excluded from the fits is that when all
of the data are plotted it became clear that only the stars falling
within the restricted ranges in equations~\ref{eqn:BB} and
\ref{eqn:CC} behaved in a manner that would make accurate calibration
possible.  Furthermore, Equations~\ref{eqn:BB} and \ref{eqn:CC} are
chosen to exclude ranges in stellar properties that some KOIs may
have, but which are not represented among our test stars.  The
parameters measured for stars outside of this range were either less
accurate or exhibited large systematic deviations from their
literature values.  In any case, it is still possible to distinguish
{\it how} the stars outside of this range differed from those inside
the range (e.g., that they were cooler or had lower \logg values).

All three of Figures~\ref{Fig:feh_trend}$-$\ref{Fig:logg_trend} show
that there are systematic trends in the differences between the
parameter values fit here and the previously-published values.  Note
that there are stars among this set measured using different methods,
but all lie along the same linear trends.  To quantify these trends,
an unbiased linear least squares fit to all of the points is found and
shown in the figures.  These fits lead to corrective relationships
that can be used to place the measured stellar properties on a scale
defined by the test stars:

\begin{equation}\label{eqn:DD}
{\rm [Fe/H](corr.) = 0.4904{\times}[Fe/H] + 0.0553}
\end{equation}

\begin{equation}\label{eqn:EE} 
{\rm T_{eff}(corr.) = 1.0953{\times}T_{eff} - 465K}
\end{equation}

\begin{equation}\label{eqn:FF}
{\rm log(g)(corr.) = 0.3489{\times}log(g) + 2.949} 
\end{equation}

\noindent
Here, the corrected value of the parameter is labelled parenthetically
with ``(corr.)''  and is expressed as a function of the uncorrected
parameter obtained during step~6 of the method described if
\S\ref{subsec:modelfitting}.  Using
equations~\ref{eqn:DD}$-$\ref{eqn:FF}, the range over which the
corrections are applicable (ie. the range over which the spectral fits
can be calibrated) can now be expressed in terms of the corrected
stellar properties:

\begin{equation}\label{eqn:GG}
{\rm
  4750K{\leq}~T_{eff}\leq7200K~\cap~log(g)\geq4.15~\cap~[Fe/H]\geq-0.25}
\end{equation}

\begin{equation}\label{eqn:HH}
{\rm
  5500K{\leq}~T_{eff}\leq7200K~\cap~log(g)\geq4.00~\cap~[Fe/H]\geq-0.25}
\end{equation}

The scatter of points around the linear fits in
Figures~\ref{Fig:feh_trend}$-$\ref{Fig:logg_trend} provides an
estimate for the uncertainties in the corrected stellar parameters.
The distribution of the points around the linear fits can be described
in terms of standard deviations, where $\sigma=0.10$~dex for \feh,
$\sigma=59$~K for \teff, and $\sigma=0.13$ for \logg.  The scatter
reflects a combination of the uncertainties from the previous
measurements and those presented here.  To determine the contribution
to the total uncertainty from the latter, one could determine an error
on each parameter that, when added in quadrature to the uncertainties
in the parameters quoted in the literature, would result in the linear
fit having $\chi^{2}_\nu=1$.  To do this, the 1-sigma errors on the
new parameter values would need to be ${\rm \sigma([Fe/H])=0.066}$,
${\rm \sigma(T_{eff})=9~K}$, and ${\rm \sigma(log(g))=0.12}$.
Evidently for \feh\ and \teff\ the uncertainties quoted for the
literature values tend to dominate the total uncertainty so that this
method could underestimate the uncertainty of the new parameter fits.
This may be the result of at least some of the uncertainties quoted in
the literature having been overestimated.  In contrast, for \logg, the
contribution of the new uncertainties to the total error is larger and
this method is useful to estimate the uncertainty.

However, since there may be unknown effects that could influence the
fits, we have chosen to adopt a more conservative uncertainty on each
measurement.  The standard deviations of data around the fitted lines
probably represents an upper limit to uncertainties within this
well-characterized range of stellar properties.  With that in mind, we
have adopted a $1\sigma$ uncertainty of 75~K for \teff, 0.10~dex for
\feh, and 0.15 for \logg.  The stellar properties of our modelled
stars are given in Table~\ref{tab:KOIdata} and referenced by KOI
number and KIC identification number.

\section{REVISED STELLAR AND PLANETARY RADII}\label{sect:radii}

In total, 368 good quality spectra were obtained of 352 stars. From
this master sample, 226 spectra for 220 stars had high enough quality,
were not now known to harbor false positive planets, and had
appropriate stellar atmospheric parameters to allow a new estimate of
stellar radius and hence new estimates of exoplanet candidate radii. A
total of 368 exoplanet candidates orbit these 220 stars. The Kepler
magnitude distribution of these 220 stars is shown in
Figure~\ref{Fig:kepmags_hist}.

To begin, the 226 stellar spectra were separated into three
\feh\ ranges: ($-0.25:-0.10$), ($-0.10:+0.20$), and
($+0.20:+0.50$). Within each \feh\ range, measured effective
temperature and surface gravity were used to estimate stellar
luminosity using the so-called Version~2 Yale-Yonsei (YY) isochrones
\citep{demarqueetal04} with solar abundance ratios (i.e. $\alpha = 0$)
and \feh~$= -0.28$, $+0.04$, and $+0.38$, respectively, as provided in
the on-line
version\footnote{{http://www.astro.yale.edu/demarque/yyiso.html}}.
Stellar luminosity in solar units was estimated for a given
(\teff,\logg) estimate by determining the median stellar luminosity in
the ranges \teff\ $\pm75$~K and \logg\ $\pm0.15$. Given the magnitude
of the (\teff,\logg) uncertainties, it was deemed appropriate to
search the on-line YY~Version~2 isochrone grid without further
interpolation.  Stellar radius in solar units was then estimated using
the standard relation

\begin{equation}\label{eqn:II}
R_{\star} = \sqrt{L_{\star}/T_{eff,\star}^4}
\end{equation}

Stellar radii uncertainties can be estimated in two ways. First,
within the search box defined by the (\teff,\logg) uncertainties, the
standard deviation of the mean luminosity $\sigma_L$ can be
computed. The radii uncertainty was estimated as follows:

\begin{equation}\label{eqn:JJ}
\sigma_R = \left (
{\sqrt{(L_{\star}+\sigma_L)/T_{eff,\star}^4} +
\sqrt{(L_{\star}-\sigma_L)/T_{eff,\star}^4}}
\right ) /2
\end{equation}

Second, 6 stars were observed at least twice and sometimes four times
on separate nights during different observing runs separated by
months.  Stellar radius uncertainty can be estimated from the
dispersion in stellar radius estimates from these individual
observations. From both methods in combination, a conservative
uncertainty for ${\rm R_\star}$ of $\pm 0.05 R_{\odot}$ is adopted.

\begin{figure}
\epsscale{1.}
\plotone{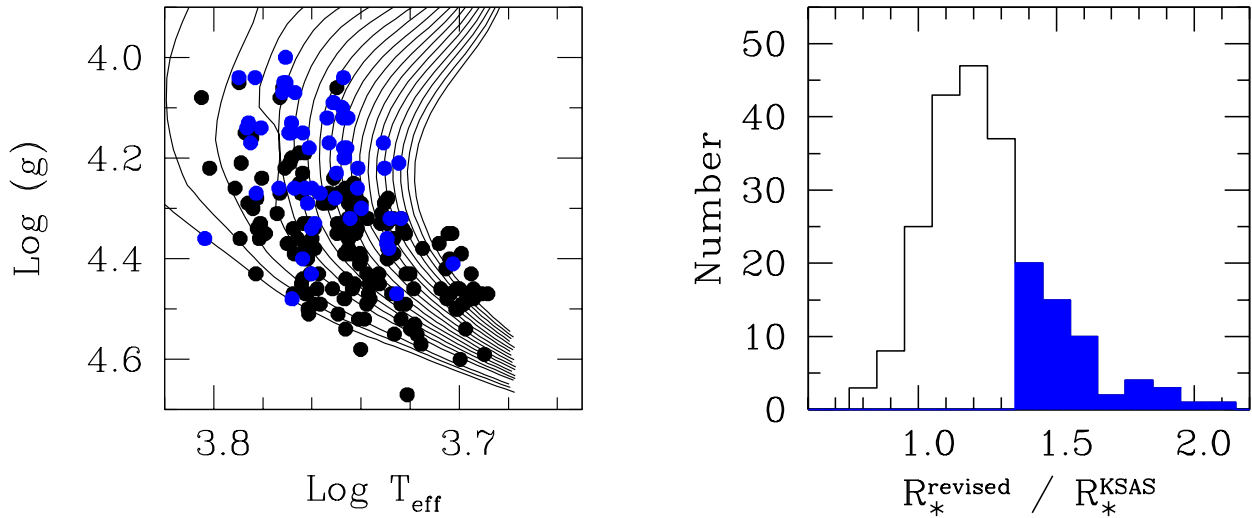}
\caption{ {\it Left panel:} The 220 stars (large black and blue
  circles) with new radii measurements compared to solar metallicity
  Version 2 Yale-Yonsei isochrones. Metal-poor and metal-rich
  isochrones were used to estimate stellar radii but are not shown
  here. Stars with blue points have revised radii $\geq1.35$
  times their KSAS radii.  {\it Right panel:} Distribution of
  $R_{\star}^{revised}$ / $R_{\star}^{KSAS}$, where stars with revised
  radii $\geq1.35$ times their KSAS radii are indicated by solid blue
  bins. Clearly, stars with larger revised radii tend to be more
  evolved than previously assumed.  (The data used to create this
  figure are available in the online journal.)
} \label{Fig:YYfits_stellar_radii}
\end{figure}

\begin{figure}
\plotone{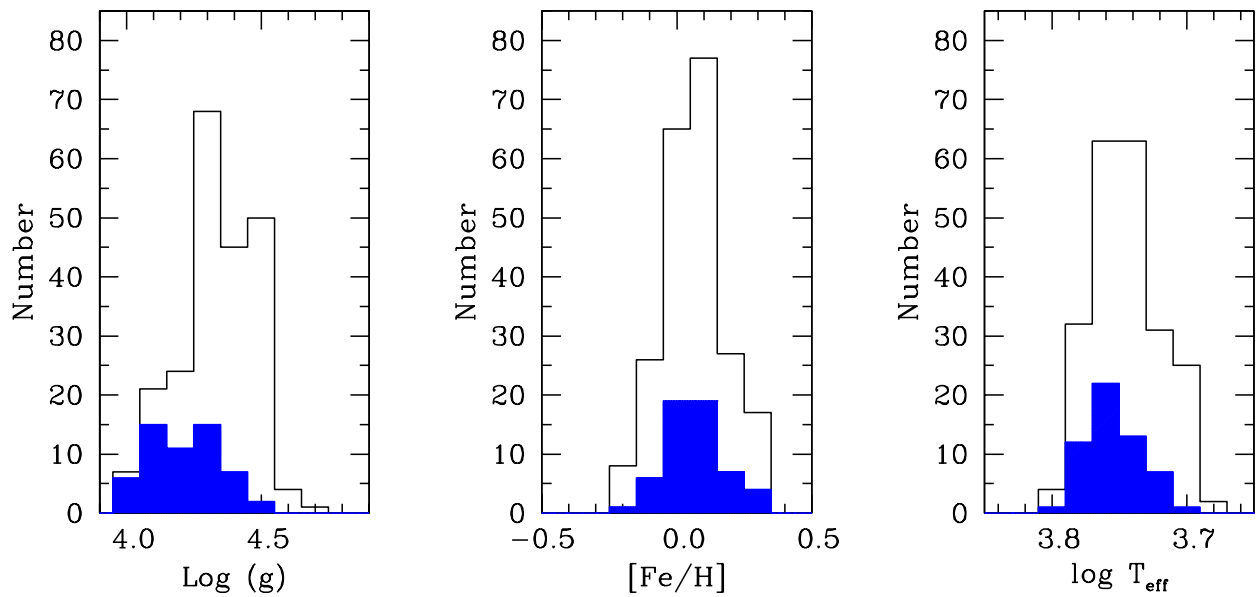}
\caption{From left-to-right, the distribution of 220 stars as a
  function of \logg, \feh, and \teff. The solid blue inserts indicate
  the sub-sample with revised radii $\geq1.35$ times their KSAS
  radii. Again, this subsample tends have lower surface gravity and
  higher temperature than the full sample.  There is no obvious
  difference in the \feh\ distribution of these two samples.  (The
  data used to create this figure are available in the online
  journal.)
} \label{Fig:YYfits_vs_parameter}
\end{figure}

The new stellar radii estimates are provided in
Table~\ref{tab:KOIdata}.  Only a portion of this long table is
presented here.  The entire table is made available in the electronic
version.  For stars observed multiple times, individual radii
estimates were averaged into a single value. How do these new
estimates compare to the best previous available radii estimates from
the Kepler Science Analysis System (KSAS)\footnote{The KSAS was a {\it
    Kepler} Mission database storing the best available estimates for
  stellar and candidate planet properties.  Stellar radii were based
  on KIC photometry for most stars in the magnitude range of interest
  here.}?  As Figure~\ref{Fig:YYfits_stellar_radii} illustrates, there
is a formal offset towards larger radii estimates.  The radii of 87\%
of these stars are revised upwards (and 13\% downwards), although some
of these revised radii are insignificant given the uncertainties in
the stellar radii.  For about 26\% (58) of the stars, the revised
radii are skewed upwards with $R_{\star}^{revised} \geq 1.35 \times
R_{\star}^{KSAS}$. As Figures~\ref{Fig:YYfits_stellar_radii} and
\ref{Fig:YYfits_vs_parameter} illustrate, these stars tend to to be
more evolved than the sample as a whole and relative to their
properties listed by KSAS.  In other words, it appears that many KIC
stars are larger than previously assumed. In turn, exoplanet
candidates orbiting those stars must be larger by the same relative
amount.

Revised exoplanet candidate radii estimates can be derived from the
revised stellar radii measurements from the simple geometric
approximation:

\begin{equation}\label{eqn:KK}
R_p^{revised} =
(R_{\star}^{revised} / R_{\star}^{KSAS}) \times R_p^{KSAS}
\end{equation}

\noindent where initial stellar and exoplanet radii come from KSAS.
Stellar radii use solar units ($R_\odot$) while exoplanet candidate
radii use Earth radius units ($R_\oplus$).  The on-line
Table~3 provides the revised exoplanet candidate
radii.

\begin{figure}
\plotone{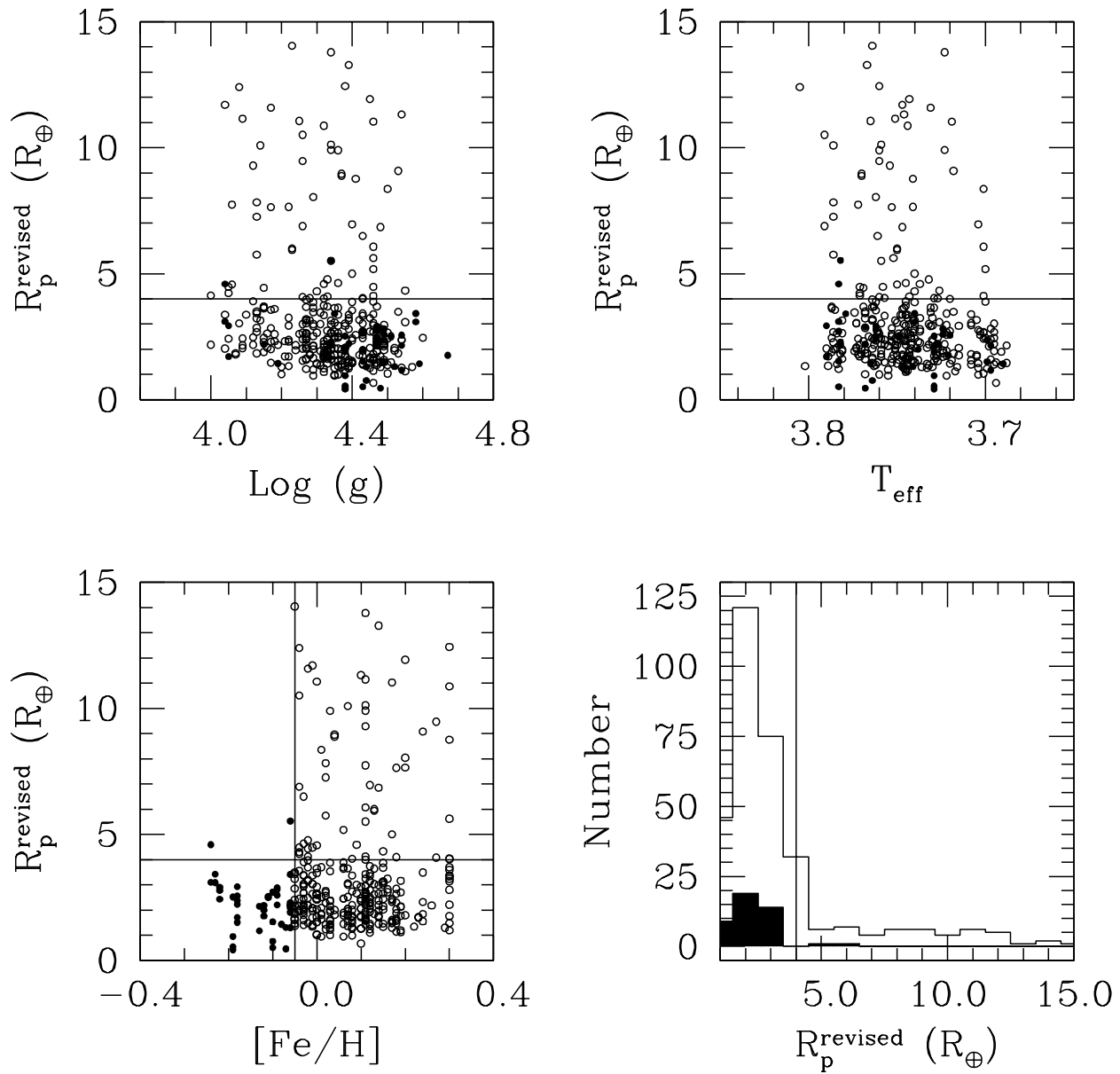}
\caption{The top panels and the lower left panel compare revised
  exoplanet candidate radii (units: $R_\oplus$) to host star
  characteristics. In all three panels, solid symbols correspond to
  exoplanet candidates associated with host stars with ${\rm [Fe/H] \leq
  -0.05}$ while open symbols correspond to stars with higher
  metallicity. Radii distributions for exoplanets orbiting lower
  metallicity stars (${\rm [Fe/H] \leq -0.05}$) (solid histogram) and solar
  or greater metallicities (open histogram) are compared in the lower
  right panel. Exoplanet candidates with $R_p^{revised} \geq 4
  R_\oplus$ are found at all temperatures and luminosities but are
  preferentially associated with stars of higher metallicity, in
  agreement with conclusions of \citet{buchhaveetal12}.  (The data
  used to create this figure are available in the online journal.)
} \label{Fig:planets_vs_parameter}
\end{figure}

Figure~\ref{Fig:planets_vs_parameter} compares revised exoplanet
candidate radii to the characteristics of their host stars. As
previously shown by \citet{buchhaveetal12} and discussed in
\S\ref{sect:discussion}, exoplanet candidates with $R_p^{revised} \geq
4 R_\oplus$ are much more likely to be associated with higher
metallicity stars, while smaller exoplanet candidates are found around
stars spanning the entire metallicity range of our sample.

\section{DISCUSSION}\label{sect:discussion}

The new stellar characteristics derived from this spectroscopic study
have refined the properties of a large sample of KOIs, revealing
statistical trends and identifying a number of individual KOIs as
excellent targets for more detailed follow-up and potential
confirmation as systems harboring small habitable zone planets.

In \S\ref{sect:radii} we found that 26\% of the KOI stars had radii
significantly larger than their values based on the initial
photometric data available to the mission.  This effect could be due
to systematic errors in the photometrically-derived stellar properties
like \logg, selection effects in the magnitude-limited KOI sample,
transit detectability dependence on stellar radius, or a combination
of factors.  In the case of these data, almost all of the stars with
radii revised upwards by a factor of 1.35 or greater have
\teff$>$5200~K, and their positions on the \logg$-$log(\teff) plot of
Figure~\ref{Fig:YYfits_stellar_radii} show that many represent a
population of relatively evolved stars compared to the other stars of
comparable effective temperature.

The lower \logg\ values measured here may be compared to those of
\citet{verneretal11} who used asteroseismic methods on the {\it
  Kepler} light curve data to determine radii for 514 solar-type stars
in the apparent magnitude range $7-12$.  For stars with
\logg$>4.0$~dex and a wide range of effective temperature, the mean
asteroseismic \logg\ values were 0.23~dex lower than those reported in
the KIC.  The corresponding stellar radii were larger as well.
Another sample of stars with asteroseismic \logg\ values was compared
to KIC \logg\ by \citet{brunttetal12}.  They found asteroseismic
\logg\ values were lower than those in the KIC by an average of
0.05~dex.  They attributed the lower mean difference with respect to
the KIC to the inclusion of stars with \logg$<4.0$ in the sample, for
which asteroseismic \logg\ values are in better agreement.
\citet{verneretal11} noted that their asteroseismic sample could be
skewed by a Malmquist bias, which would preferentially select more
evolved and intrinsically brighter stars, as well as by the improved
detectability of the higher amplitude oscillations associated with
stars of lower \logg.  In the case of the spectroscopically analyzed
sample presented here, the Malmquist bias would be in effect along
with the counteracting bias favoring detectability of transits across
smaller stars.  These two biases were examined by
\citet{gaidosandmann13} who predicted that the Malmquist bias would
have the dominant effect, and the transit sample should be relatively
overabundant in large stars compared to stars at the same temperature
and apparent brightness.  In addition to biases in the KOI sample as a
whole, this spectroscopic sample was constructed to include many of
the (relatively rare) smallest planet candidates for follow-up, a
choice that may also select KOI stars with anomalous radii.  It is
clear that a full understanding of these biases is necessary to get
better estimates for planet occurrence rates and that large,
spectroscopic samples like the one presented here will play an
important role.  A similar spectroscopic study of ``control'' stars,
perhaps {\it Kepler} stars showing no transits, may be of merit as
well.

The revised values for stellar and planet radii have some implications
for the mission goal to determine the frequency of Earth-sized planets
orbiting Sun-like stars in a habitable zone.  The radii of some
planets must be revised significantly upwards, perhaps pushing them
outside the size range likely for rocky Earth-like bodies.  An
additional effect is that higher luminosities implied by an increase
in stellar radius move the habitable zones for these stars outwards
from the star.  As a consequence, the orbital periods of habitable
zone planets must be longer.

Despite the apparent decrease in the number of small planets, these
spectra provide additional evidence to favor certain candidates as
among the most interesting targets for the goals of the {\it Kepler}
Mission.  An example candidate host star is KOI2931, the star shown in
Figures~\ref{Fig:K02931spectra} and \ref{Fig:K02931interp}.  KOI2931
hosts a single known planet candidate, KOI2931.01, with an orbital
period of 99.248~days.  With new stellar properties \teff$=4991$~K,
\logg$=4.49$, \feh$=-0.03$ and ${\rm R_\star=0.85R_\odot}$, the planet
radius of KOI2931.01 is estimated to be 2.1~$R_\oplus$.  The isochrone
fit for this star corresponds to a stellar mass of ${0.78M_\odot}$ and
a planet equilibrium temperature of 326~K is found assuming an albedo
of 0.3 and a circular orbit.  KOI2931.01 is one example of a good
candidate for a super-Earth orbiting in the habitable zone.

A correlation between the incidence of relatively large planet
candidates and relatively high host star metallicity (selecting large
planets at ${\rm R_p=4.0R_\oplus}$) was previously seen
spectroscopically in a smaller sample of brighter KOI stars by
\citet{buchhaveetal12}.  The KOI stars in their sample were almost all
brighter than ${\rm m_{kep}=14}$, but our and their data sets overlap
in apparent brightness.

There are various ways to examine the significance of the apparent
deficit of large planet candidates around low metallicity host stars
(\feh$<-0.05$) in this sample.  First, note that 5 planet candidates
in this sample (225.01, 998.01, 1067.01, 1226.01 and 1483.01) are all
too large (${\rm >40R_\oplus}$) while the remainder are reasonable
sizes for planets (${\rm <20R_\oplus}$).  These 5 objects are
considered likely false positives and excluded from further
consideration.  This results in 46 candidate planets orbiting host
stars of \feh$<-0.05$ and 317 orbiting host stars of \feh$\geq-0.05$.
A K-S test comparing the planet size distributions of the these two
samples reveals a difference with a confidence level of 98\%.  As a
second test, random subsamples of 46 candidate planets are drawn from
the sample of 317 candidates orbiting host stars with \feh$\geq-0.05$
and compared to the 46 candidate planets orbiting lower metallicity
stars.  A set of 1 million random subsamples reveals that the most
probable number of large planet candidates (${\rm R_p>4R_\oplus}$)
orbiting high metallicity host stars is 8 or 9, and that 2 or fewer
large planet candidates occur just 0.4\% of the time (2 is the number
of large planet candidates orbiting the low metallicity host stars).

A fraction of the candidate planet sample may be false positives and
this effect is considered next.  Note that 38 known or likely false
positives have already been removed from the sample of 352 stars as
part of creating the candidate planet sample, but others likely
remain.  Also, 243 out of 363 planet candidates are members of
multi-planet systems and these have a very high likelihood of being
true planets rather than eclipsing binary stars
\citep{lissaueretal12}.  However, multi-planet systems may still be
considered false positives in the sense that their planet radii can be
underestimated due to host star blending \citep{fressinetal13}.  A
detailed treatment might be useful to simulate the effects of false
positives in the sample, but a simpler approach is taken here.  If a
liberal reduction is made to the sample size in an effort to simulate
the removal of false positives, it will weaken inferences drawn from
these data.  \citet{fressinetal13} predict false positive rates in
five planet size ranges: 17.7\% for ${\rm R_p=6-22R_\oplus}$, 15.9\%
for ${\rm R_p=4-6R_\oplus}$, 6.7\% for ${\rm R_p=2-4R_\oplus}$, 8.8\%
for ${\rm R_p=1.25-2R_\oplus}$ and 12.3\% for ${\rm
  R_p=0.8-1.25R_\oplus}$.  When individual planets are removed from
our sample at these rates, the K-S test signficance of the differences
in planet size distribution on host star metallicity drops to 96\%.
The test of selecting random samples of high metallicity stars to
match the sample size of the low metallicity stars reveals that 2 or
fewer large planet candidates occur around high metallicity stars
1.7\% of the time.

The tests show a dependence between host star metallicity and the
occurance rate of large transiting planets, much like for the sample
of \citet{buchhaveetal12}.  It is not surprising to see a similar
pattern in these data, but the fainter stars analyzed here probe a
significantly larger volume of space, showing that these effects
persist across the different stellar populations.  The apparent
threshold value of metallicity is chosen at \feh$=-0.05$ to match the
appearance of the lower left panel in
Figure~\ref{Fig:planets_vs_parameter}, but the discrete and relatively
sparse set of model spectra used to determine \feh\ may slightly
distort this plot.  The lines drawn at ${\rm R_p=4.0R_\oplus}$ were
also chosen by eye, but could have as well been taken at a somewhat
smaller radius (ie. at ${\rm R_p=3-4R_\oplus}$).  There are no obvious
trends in the incidence of planet candidates with ${\rm
  R_p>4R_\oplus}$ with respect to either \logg\ or \teff.  Similarly,
no dependence was found between metallicity and the number of planets
detected around the KOI stars.  Given the lack of large planets
detected (in short period orbits) around low metallicity host stars,
the efficiency of planet migration may be dependent on metallicity, or
perhaps large planets simply cannot form around such stars at any
orbital distance.

\section{DATA AVAILABILITY}\label{sect:dataavailable}

The reduced spectra and products from our model fits are made
available on the CFOP website${\rm ^5}$.  The CFOP website organizes
data for each KOI and confirmed {\it Kepler} exoplanets, including the
products of many follow-up observations.  The data products
contributed from this spectroscopy program include the reduced
spectrum data files, stellar properties, plots of the spectra, fitted
synthetic models plotted alongside the observed spectra (similar to
Figure~\ref{Fig:K02931spectra}) and plots similar to
Figure~\ref{Fig:K02931interp} showing the interpolation between
gridpoint fit values.  Additional follow-up spectra and their fits
will be added in the future.

\section{SUMMARY}\label{sect:conclusions}

A spectroscopic analysis of a large sample of stars known as {\it
  Kepler} Objects of Interest (KOIs) is presented.  In the case of
most of these KOIs, the stellar characterization, and by extension
candidate planet properties, had been based on broadband photometry
available from the pre-launch {\it Kepler} Input Catalog survey.
Spectral follow-up, like that presented here, proves important to
improve the accuracy of the KOI stellar properties, identify
interesting individual planet systems and perform accurate statistical
studies of the KOI list as a whole.  The results of model spectra fits
(values for \teff, \logg, and \feh) are given for 268 stars.
Isochrone fits are used to provide revised radii for 220 KOI stars and
their 368 planets.  The spectra and results from this survey are made
available to the public through the online CFOP archive.

The spectral and isochrone fits reveal that many of the KOI stars have
larger radii than previously assumed.  About 26\% of the stars for
which new radii were determined require corrections to their assumed
radii of a factor of 1.35 or greater, and the isochrone fits for 87\% of
the stars suggest some increase in radius.  The stars requiring the
largest upward adjustment in radius represent a relatively evolved
subset of the sample.  The increases in stellar radii also require a
reevaluation of the radii derived for the planet candidates hosted by
these stars.  The planet radii need to be scaled upwards by
approximately the same ratio as their host stars.

Despite the fact that the revised planet radii are overall larger than
previously assumed, there are candidate planets in this sample that
are now better vetted and continue to be likely small planets in the
habitable zone of Sun-like stars.  The example of KOI2931 is presented
as a good candidate for a super-Earth planet orbiting in the habitable
zone of a 4991~K dwarf.

The frequency of large KOI planets in the sample depends on host star
metallicity in a manner similar to that found for a sample of brighter
KOI stars by \citet{buchhaveetal12}.  The fainter, larger sample of
$4750-7000$~K dwarf KOIs analyzed in our program shows that these
results extend through a larger volume of space and that the
occurrence of large planets (${\rm R_p>3-4R_\oplus}$) depends on a
threshold metallicity near \feh$=-0.05$.  The large planet
candidates are found almost exclusively around stars with metallicity
higher than this value.  In contrast, small planet candidates are found
around stars spanning the full metallicity range examined in this
study.

\acknowledgments
Our work was made possible through the efforts of many others.  Among
them are those in the {\it Kepler} Science Office and science team.
At the telescope we always received excellent help from our observing
assistants, and help from additional observers Jay Holberg, Ken
Mighell, and Jason Rowe.  Codes used in our modelling were adopted
from work by Greg Doppmann and we received help to compile a list of
properties for our test star sample from Lars Buchhave and Thomas
Kallinger.  We also wish to thank the referee for helpful
  suggestions that were incorporated into this work.  Financial
support for the work was provided by the NASA {\it Kepler} Mission and
Cooperative Agreement AST--0950945 to NOAO.

Facilities: \facility{Mayall}, \facility{Kepler}

\newpage

\pagestyle{empty}

\begin{figure}
\epsscale{0.55}
\plotone{Figure1.eps}
\caption{Number of KOI stars vs. Kepler magnitude. The unfilled
  histogram represents the entire KOI sample as of September 2012,
  i.e. this was the parent sample for the project described in this
  paper. The grey-filled histogram represents the total observed
  sample (see Table~\ref{tab:KOIdata}).  81\% of these stars have
  $m_{Kep}>14$. The black-filled histogram represents the sub-sample
  of observed stars with new radii estimates (also see
  Table~\ref{tab:KOIdata}).
} \label{Fig:kepmags_hist}
\end{figure}

\clearpage

\clearpage

\begin{figure}
\epsscale{0.55}
\plotone{Figure5.eps}
\caption{A comparison of preliminary interpolated \feh\ values found
  for 24 test stars of main sequence luminosity class to
  the values previously reported in the literature.  The abscissa
  values are the preliminary interpolated \feh\ while the ordinate
  shows the difference between these values and those from the
  literature.  Error bars represent the uncertainties quoted for the
  \feh\ values in the literature.  The straight line shows an unbiased
  least-squares fit through the points and defines a correction to be
  made to remove systematic errors in the preliminary \feh\ estimates.
  This correction results in the final adopted \feh\ estimates.  The
  scatter around the fit provides an estimate for the \feh\
  uncertainties.  On this plot, open circles represent HAT Project
  exoplanet host stars, the filled triangle is the relatively hot
  dwarf from \citet{luckheiter07}, star-shaped symbols are various KOI
  stars, and the diamonds represent stars from
  \citet{valentifischer05}.
} \label{Fig:feh_trend}
\end{figure}

\begin{figure}
\epsscale{0.55}
\plotone{Figure6.eps}
\caption{A comparison of preliminary interpolated \teff\ values found
  for 24 test stars of main sequence luminosity class to
  the values previously reported in the literature.  The abscissa
  values are the preliminary interpolated \teff\ while the ordinate
  shows the difference between these values and those from the
  literature.  Error bars represent the uncertainties quoted for the
  \teff\ values in the literature.  The straight line shows an unbiased
  least-squares fit through the points and defines a correction to be
  made to remove systematic errors in the preliminary \teff\ estimates.
  This correction results in the final adopted \teff\ estimates.  The
  scatter around the fit provides an estimate for the
  \teff\ uncertainties.  See Figure~\ref{Fig:feh_trend} for an
  explanation of the plotting symbols.
} \label{Fig:teff_trend}
\end{figure}

\begin{figure}
\epsscale{0.55}
\plotone{Figure7.eps}
\caption{A comparison of preliminary interpolated \logg\ values found
  for 24 test stars of main sequence luminosity class to
  the values previously reported in the literature.  The abscissa
  values are the preliminary interpolated \logg\ while the ordinate
  shows the difference between these values and those from the
  literature.  Error bars represent the uncertainties quoted for the
  \logg\ values in the literature.  The straight line shows an unbiased
  least-squares fit through the points and defines a correction to be
  made to remove systematic errors in the preliminary \logg\ estimates.
  This correction results in the final adopted \logg\ estimates.  The
  scatter around the fit provides an estimate for the
  \logg\ uncertainties.  See Figure~\ref{Fig:feh_trend} for an
  explanation of the plotting symbols.
} \label{Fig:logg_trend}
\end{figure}

\newpage

\begin{deluxetable}{lclrllrlcllrcrrrcl}
\tabletypesize{\footnotesize}
\rotate
\tablecolumns{14}
\tablewidth{0pt}
\tablecaption{Test star spectral fits}
\tablehead{  && \multicolumn{6}{c}{Stellar properties from the literature} && \multicolumn{3}{c}{Values from this work} && \multicolumn{3}{c}{Difference in values} && \\
\cline{3-8} \cline{10-12} \cline{14-16}
\colhead{ID} & \colhead{} & \colhead{${\rm T_{eff}}$} & \colhead{err} & \colhead{log(g)} & \colhead{err} & \colhead{[Fe/H]} & \colhead{err} & \colhead{} & \colhead{${\rm T_{eff}}$} & \colhead{log(g)} & \colhead{[Fe/H]} & \colhead{} & \colhead{${\rm \Delta T_{eff}}$} & \colhead{${\rm \Delta log(g)}$} & \colhead{${\rm \Delta [Fe/H]}$} & \colhead{} & \colhead{reference} \\
}
\startdata
BD+05 3640 && 5104 & 44 & 4.85 & 0.06 & -1.14 & 0.03 && 5079 & 4.01 & -1.34 && -25 & -0.842 & -0.196 && \citet{valentifischer05} \\
HAT-P-2\tablenotemark{a} && 6290 & 110 & 4.21 & 0.09 & 0.12 & 0.08 && 6274 & 3.77 & 0.04 && -16 & -0.440 & -0.080 && \citet{bakosetal07} \\
HAT-P-3\tablenotemark{a} && 5185 & 46 & 4.58 & 0.04 & 0.27 & 0.04 && 5072 & 3.95 & 0.49 && -113 & -0.630 & 0.220 && \citet{torresetal07} \\
HAT-P-4\tablenotemark{a} && 5860 & 80 & 4.14 & 0.04 & 0.24 & 0.08 && 5787 & 3.42 & 0.25 && -73 & -0.720 & 0.010 && \citet{kovacsetal07} \\
HAT-P-6\tablenotemark{a} && 6570 & 80 & 4.22 & 0.03 & -0.13 & 0.08 && 6487 & 3.97 & -0.30 && -83 & -0.250 & -0.170 && \citet{noyesetal08} \\
HAT-P-8\tablenotemark{a} && 6200 & 80 & 4.15 & 0.03 & 0.01 & 0.08 && 6073 & 3.25 & -0.10 && -127 & -0.900 & -0.110 && \citet{lathametal09} \\
HAT-P-10\tablenotemark{a} && 4980 & 60 & 4.56 & 0.02 & 0.13 & 0.08 && 4954 & 4.58 & 0.17 && -26 & 0.020 & 0.040 && \citet{bakosetal09a} \\
HAT-P-12 && 4591 & 60 & 4.61 & 0.01 & -0.36 & 0.04 && 4391 & 4.21 & -0.36 && -200 & -0.400 & 0.000 && \citet{hartmanetal09} \\
\enddata
\tablenotetext{a}{Star is one of 24 used to calibrate the model fits (see \S\ref{subsec:teststarfits})}
\label{tab:teststars}
\end{deluxetable}

\newpage

\begin{deluxetable}{rrrccrcl}
\tabletypesize{\scriptsize}
\tablecolumns{8}
\tablewidth{0pt}
\tablecaption{Stellar properties determined for KOIs}
\tablehead{ 
\colhead{KOI} & \colhead{Kepler ID} & \colhead{${\rm m_{Kep}}$} & 
\colhead{${\rm T_{eff}\tablenotemark{1}}$} & \colhead{${\rm log(g)}$\tablenotemark{2}} & 
\colhead{${\rm [Fe/H]}$\tablenotemark{3}} & \colhead{R$_\star$\tablenotemark{4}} & notes \\
\colhead{} & \colhead{} & \colhead{(mag.)} & 
\colhead{(K)} & \colhead{(cgs)} & 
\colhead{} & \colhead{(R$_\odot$)} & \colhead{}
}
\startdata
13 & 9941662 & 9.958 & \nodata & \nodata & \nodata & \nodata & b \\
13 & 9941662 & 9.958 & \nodata & \nodata & \nodata & \nodata & b \\
14 & 7684873 & 10.470 & \nodata & \nodata & \nodata & \nodata & bg \\
14 & 7684873 & 10.470 & \nodata & \nodata & \nodata & \nodata & bg \\
70 & 6850504 & 12.498 & 5562 & 4.32 & 0.12 & 1.13 &  \\
74 & 6889235 & 10.960 & \nodata & \nodata & \nodata & \nodata & bg \\
115 & 9579641 & 12.791 & 6065 & 4.43 & -0.10 & 1.08 &  \\
136 & 7601633 & 13.439 & 6035 & 4.36 & 0.01 & \nodata & g \\
\enddata
\tablenotetext{1}{Uncertainty in \teff\ is $\pm75$~K}
\tablenotetext{2}{Uncertainty in \logg\ is $\pm0.15$}
\tablenotetext{3}{Uncertainty in \feh\ is $\pm0.10$}
\tablenotetext{4}{Uncertainty in R$_\star$ is $\pm0.05$~R$_\odot$}
\tablecomments{
 Values in the notes column flag the following conditions: 
(a) ${\rm T_{eff}<4750K}$; (b) ${\rm T_{eff}>7200K}$; 
(c) ${\rm log(g)<4.00}$ and ${\rm 5500K{\leq}T_{eff}{\leq}7200K}$; 
(d) ${\rm log(g)<4.15}$ and ${\rm 4750K{\leq}T_{eff}<5500K}$; 
(e) ${\rm [Fe/H]<-0.25}$; 
(f) model fit reached a parameters limit 
(g) Now known to be a false positive. 
}
\label{tab:KOIdata}
\end{deluxetable}

\newpage

\begin{deluxetable}{lrllrlll}
\tabletypesize{\scriptsize}
\tablecolumns{7}
\tablewidth{0pt}
\tablecaption{Data used in Figures \ref{Fig:YYfits_stellar_radii}--\ref{Fig:planets_vs_parameter}}
\tablehead{ 
\colhead{Planet ID} & \colhead{Parent star} & \colhead{log(${\rm T_{eff}}$)} & \colhead{${\rm log(g)}$} & 
\colhead{${\rm [Fe/H]}$} & \colhead{${\rm R_\star}$} & 
\colhead{${\rm R_\star^{revised}/R_\star^{KSAS}}$} & \colhead{${\rm R_p}$} \\
\colhead{} & \colhead{(KIC ID)} & \colhead{(K)} & \colhead{(cgs)} & 
\colhead{} & \colhead{(${\rm R_\odot}$)} & 
\colhead{} & \colhead{(R$_\oplus$)} 
}
\startdata
K00070.01 & 6850504 & 3.745 & 4.32 & 0.12 & 1.13 & 1.21 & 3.74 \\
K00070.02 & 6850504 & 3.745 & 4.32 & 0.12 & 1.13 & 1.21 & 2.32 \\
K00070.03 & 6850504 & 3.745 & 4.32 & 0.12 & 1.13 & 1.21 & 3.35 \\
K00070.04 & 6850504 & 3.745 & 4.32 & 0.12 & 1.13 & 1.21 & 1.10 \\
K00070.05 & 6850504 & 3.745 & 4.32 & 0.12 & 1.13 & 1.21 & 1.24 \\
K00115.01 & 9579641 & 3.783 & 4.43 & -0.10 & 1.08 & 0.81 & 2.71 \\
K00115.02 & 9579641 & 3.783 & 4.43 & -0.10 & 1.08 & 0.81 & 1.53 \\
K00115.03 & 9579641 & 3.783 & 4.43 & -0.10 & 1.08 & 0.81 & 0.51 \\
\enddata
\label{tab:planetinfo}
\end{deluxetable}


\newpage

\begin{thebibliography}{}

\bibitem[Bakos et al.(2002)]{bakosetal02} Bakos, G. \'A., L\'az\'ar,
  J., Papp, I., S\'ari, P. \& Green, E. M. 2002, \pasp, 114, 974

\bibitem[Bakos et al.(2007)]{bakosetal07} Bakos, G. \'A. et al. 2007,
  \apj, 670, 826B

\bibitem[Bakos et al.(2009a)]{bakosetal09a} Bakos, G. \'A. et
  al. 2009a, \apj, 696, 1950B

\bibitem[Bakos et al.(2009b)]{bakosetal09b} Bakos, G. \'A. et
  al. 2009b, \apj, 707, 446B

\bibitem[Bakos et al.(2011)]{bakosetal11} Bakos, G. \'A. et al. 2011,
  \apj, 742, 116B

\bibitem[Barclay et al.(2012)]{barclayetal12} Barclay, T. et al. 2012,
  in prep.

\bibitem[Batalha et al.(2010)]{batalhaetal10} Batalha, N. M. et
  al. 2010, \apj, 713L, 103B

\bibitem[B\'eky et al.(2011)]{bekyetal11} B\'eky, B. et al. 2011,
  \apj, 734, 109B

\bibitem[Borucki et al.(2010)]{boruckietal10} Borucki, W. J. et al.
  2010, Science, 327, 977

\bibitem[Brown et al.(2011)]{brownetal11} Brown, T. M., Latham, D. W.,
  Everett, M. E., \& Esquerdo, G. A. 2011, \aj, 142, 112

\bibitem[Bruntt et al.(2012)]{brunttetal12} Bruntt, H. et al. 2012,
  \mnras, 423, 122

\bibitem[Buchhave et al.(2010)]{buchhaveetal10} Buchhave, L. A. et
  al. 2010, \apj, 720, 1118B

\bibitem[Buchhave et al.(2011)]{buchhaveetal11} Buchhave, L. A. et
  al. 2011, \apj, 733, 116B

\bibitem[Buchhave et al.(2012)]{buchhaveetal12} Buchhave, L. A. et
  al. 2012, Nature, 486, 375B

\bibitem[Castelli \& Kurucz(2003)]{castellikurucz03} Castelli, F. \&
  Kurucz, R. L. 2003, in Proc. of the 210th Symposium of the IAU at
  Uppsala University, Uppsala, Sweden, 17-21 June, 2002. ed. by
  N. Piskunov, W. W. Weiss, \& D. F. Gray. Published on behalf of the
  IAU by the Astronomical Society of the Pacific, A20

\bibitem[Coelho et al.(2005)]{coelhoetal05} Coelho, P., Barbuy, B.,
  Mel\'endez, J., Schiavon, R. P., Castilho, B. V. 2005, \aap, 443,
  735

\bibitem[Col{\'o}n et al.(2012)]{colonetal12} Col{\'o}n, K. D.,
  Ford, E. B. \& Morehead, R. C. 2012, \mnras, 426, 342

\bibitem[Demarque et al.(2004)]{demarqueetal04} Demarque, P., Woo,
  J.-H., Kim, Y.-C., Yi, S.~K., 2004, \apjs, 155, 667D

\bibitem[Doyle et al.(2011)]{doyleetal11} Doyle, L. R. 2011, Science,
  333, 1602D

\bibitem[Fischer \& Valenti(2005)]{fischervalenti05} Fischer, D. A. \&
  Valenti, J. 2005, \apj, 622, 1102

\bibitem[Fressin et al.(2013)]{fressinetal13} Fressin, F. et al. 2013,
  \apj, 766, 81

\bibitem[Gaidos \& Mann(2013)]{gaidosandmann13} Gaidos, E. \& Mann, A. W.
  2013, \apj, 762, 41G

\bibitem[Gautier et al.(2010)]{gautieretal10} Gautier, T. N. et al.
  2010, arXiv 1001.0352

\bibitem[Grevesse \& Sauval(1998)]{grevesseandsauval98} Grevesse, N. \&
  Sauval, A. J. 1998, Space Sci. Rev., 85, 161

\bibitem[Gustafsson et al.(2003)]{gustafssonetal03} Gustafsson, B.,
  Edvardsson, B. Eriksson, K., Mizuno-Wiedner, M., J{\o}rgensen,
  U. G. \& Plez, B. 2003, in ASP Conf. Proceedings Vol. 288, Stellar
  Atmosphere Modeling, ed. I. Hubeny, D. Mihalas \& K. Werner (San
  Francisco: ASP), 331

\bibitem[Hartman et al.(2009)]{hartmanetal09} Hartman, J. D. et al.
  2009, \apj, 706, 785

\bibitem[Hartman et al.(2011a)]{hartmanetal11a} Hartman, J. D. et al.
  2011a, \apj, 726, 52H

\bibitem[Hartman et al.(2011b)]{hartmanetal11b} Hartman, J. D. et al.
  2011b, \apj, 728, 138H

\bibitem[Holman et al.(2010)]{holmanetal10} Holman, M. J. et al.
  2010, Science, 330, 51H

\bibitem[Howell et al.(2012)]{howelletal12} Howell, S. B. et al.
  2012, \apj, 746, 123H

\bibitem[Kallinger et al.(2010)]{kallingeretal10} Kallinger, T. et al.
  2010, \aap, 522, A1

\bibitem[Kallinger et al.(2012)]{kallingeretal12} Kallinger, T. et al.
  2012, \aap, 541, 51K

\bibitem[Kovacs et al.(2007)]{kovacsetal07} Kov\'acs, G. et al. 2007,
  \apj, 670L, 41K

\bibitem[Kurucz \& Avrett(1981)]{kuruczavrett81} Kurucz, R. L. \&
  Avrett, E. H. 1981, SAOSR, 391

\bibitem[Kurucz (1992)]{kurucz92} Kurucz, R. L. 1992, in Proceedings
  of the 149th Symposium of the International Astronomical Union, The
  Stellar Populations of Galaxies, ed. B. Barbuy \& A. Renzini
  (Dordrecht: Kluwer), 225

\bibitem[Latham et al.(2009)]{lathametal09} Latham, D. W. et al. 2009,
  \apj, 704, 1107L

\bibitem[Lissauer et al.(2012)]{lissaueretal12} Lissauer, J. L. et al.
  2012, \apj, 750, 112L

\bibitem[Luck \& Heiter(2007)]{luckheiter07} Luck, R. E., \& Heiter,
  U.  2007, \aj, 133, 2464

\bibitem[Massey et al.(1988)]{masseyetal88} Massey, P., Strobel, K.,
  Barnes, J. V. \& Anderson, E. 1988, \apj, 328, 315

\bibitem[Morton(2012)]{morton12} Morton, T. D. 2012, \apj, 761, 6

\bibitem[Morton \& Johnson(2011)]{mortonjohnson11} Morton, T. D. \&
  Johnson, J. A. 2011, \apj, 738, 170

\bibitem[Noyes et al.(2008)]{noyesetal08} Noyes, R. W. et al. 2008,
  \apj, 673L, 79N

\bibitem[Quinn et al.(2012)]{quinnetal12} Quinn, S. N. et al. 2012,
  \apj, 745, 80Q

\bibitem[Santerne et al.(2012)]{santerneetal12} Santerne, A. et al. 2012,
  \aap, 545, 76

\bibitem[Sneden (1973)]{sneden73} Sneden, C. A. 1973, PhD Thesis,
  Univ. of Texas, Austin

\bibitem[Sozzetti et al.(2007)]{sozzettietal07} Sozzetti, A., Torres,
  G., Charbonneau, D., Latham, D. W., Holman, M. J., Winn, J. N.,
  Laird, J. B., O'Donovan, F. T. 2007, \apj, 664, 1190

\bibitem[Stone(1977)]{stone77} Stone, R. P. S. 1977, \apj, 218, 767

\bibitem[Torres et al.(2007)]{torresetal07} Torres, G. et al. 2007,
  \apj, 666, L121

\bibitem[Torres et al.(2010)]{torresetal10} Torres, G. et al. 2010,
  \apj, 715, 458T

\bibitem[Valenti \& Fischer(2005)]{valentifischer05} Valenti, J. A. \&
  Fischer, D. A. 2005, \apjs, 159, 141V

\bibitem[Valenti \& Piskunov(1996)]{valentipiskunov96} Valenti,
  J. A. \& Piskunov, N. 1996, \aaps, 118, 595

\bibitem[Verner et al.(2011)]{verneretal11} Verner, G. A. et al. 2011,
  \apjl, 738, L28

\bibitem[Weck et al.(2003)]{wecketal03} Weck, P. F., Schweitzer, A.,
  Stancil, P. C., Hauschildt, P. H. \& Kirby, K. 2003, \apj, 582, 1059

\end{thebibliography}
\end{document}